\documentclass[pra,onecolumn]{revtex4}
\usepackage{bbm}
\usepackage{amsfonts}
\usepackage{amsmath}
\usepackage{amssymb}
\usepackage{graphicx}

\input{tcilatex}

\begin{document}

\title{Experimental entanglement quantification and verification via
uncertainty relations}
\author{Zhi-Wei Wang }
\email{sdzzwzw@mail.ustc.edu.cn}
\author{Yun-Feng Huang}
\email{hyf@ustc.edu.cn}
\author{Xi-Feng Ren}
\author{Yong-Sheng Zhang}
\author{Guang-Can Guo}
\affiliation{Key Laboratory of Quantum Information, University of Science and Technology
of China, CAS, Hefei 230026, People's Republic of China}

\begin{abstract}
We report on experimental studies on entanglement quantification and\
verification based on uncertainty relations for systems consisting of two
qubits. The new proposed measure is shown to be invariant under local
unitary transformations, by which entanglement quantification is implemented
for two-qubit pure states. The nonlocal uncertainty relations for two-qubit
pure states are also used for entanglement verification which serves as a
basic proposition and promise to be a good choice for verification of
multipartite entanglement.

PACS number(s): 03.67.Mn, 42.50.Dv, 42.65.Lm
\end{abstract}

\maketitle

Entanglement plays a key role in quantum information processing, such as
quantum teleportation \cite{1}, efficient quantum computation \cite{2} and
entangled-assisted quantum cryptography \cite{3}. Since more and more
experimental realization of entanglement sources become available \cite{4},
it is necessary to develop efficient methods of testing the entanglement
produced by these sources. Bell inequalities and entanglement witness \cite%
{witness}, are the main tools to detect entanglement, but the construction
of witnesses for entanglement detection and quantification of\ general
states is still a highly nontrivial task \cite{other,5,add}. Some other ways
of direct detection of quantum entanglement has been proposed in Ref. \cite%
{6,Fl} and demonstrated by linear optics in Ref. \cite{wang,wal}.

During the last few years, local uncertainty relations (LURs) have been
suggested which use sums of variance of local observables to probe the
existence of entanglement in $N$-level systems \cite{hofmann}. It is easy to
implement experimentally, but unfortunately no known LUR can detect all
entangled two-qubit states and, in general it do not give quantitative
measure of entanglement. Khan and Howell \cite{howell} investigated the
method for entangled photon pairs emitted from a down-conversion source.
They perform entanglement verification using basic LURs inequalities\ and
show these inequalities have more sensitivities than a Bell's measurement
while each requiring less measurements than a Bell's measurement to obtain.\
But these LURs are not able to quantify entanglement except some special
states. Samuelsson and Bj\"{o}rk \cite{samuel} introduce the general theory
of LURs with improved characteristics and point out that the extended use of
LURs requires that the behavior of them obey certain criteria, such as
invariance under local unitary transformations (ILUT). In the following,
Kothe and Bj\"{o}rk \cite{Kothe} present a significantly improved measure
for entanglement quantification of two-qubit system with the character of
ILUT. A generalization of LUR theory, by considering nonlocal observables,
can also\ be applied to entanglement verification for\ bipartite systems and
for\ multipartite systems, they will have more applications \cite{Guhne}.

For two systems A and B, one can choose two sets of observables $\{\hat{A}%
_{i}\}$ and$\ \{\hat{B}_{i}\}$, acting on system A and B, respectively. The
local variances are given by $\delta ^{2}\hat{A}_{i}\equiv \langle \hat{A}%
_{i}^{2}\rangle -\langle \hat{A}_{i}\rangle ^{2}$, and similar for $\delta
^{2}\hat{B}_{i}$. The sums of the local variances $\sum_{i}\delta ^{2}\hat{A}%
_{i}$ and $\sum_{i}\delta ^{2}\hat{B}_{i}$, will each have a minimum lower
bound, $U_{A}$ and $U_{B}$, respectively. The local uncertainty relation%
\begin{equation}
\sum_{i}\delta ^{2}\left( \hat{A}_{i}+\hat{B}_{i}\right) \geq U_{A}+U_{B}
\end{equation}%
holds for all mixtures of product states \cite{hofmann}.

Define the covariance term as
\begin{equation}
C(\hat{A}_{i},\hat{B}_{i})=\langle \hat{A}_{i}\hat{B}_{i}\rangle -\langle
\hat{A}_{i}\rangle \langle \hat{B}_{i}\rangle \text{.}
\end{equation}%
If the LUR is to reveal entanglement, at least one of the covariance terms
in Eq. $(2)$ has to be less than zero for entangled states.

In order to quantify entanglement of a state, the meaure used must obey the
criteria of ILUT. However, the single covariance can not satisfy this
criteria. Kothe and Bj\"{o}rk \cite{Kothe} propose a new measure
\begin{equation}
G=\sum_{i,j=1}^{3}C^{2}(\sigma _{i},\sigma _{j})
\end{equation}%
which combines several covariances\ as a quantification of entanglement. It
allows entanglement quantification of all pure states and a certain range of
mixed states. $G$ has the symmetry about the Pauli operators $\sigma _{i}$ $%
(i=1,2,3)$, therefore it has the potential probability of ILUT. ILUT also
means that \textit{a shared spatial reference }is no longer needed and will
make it convenient for both theoretical and experimental investigation.

For pure states, it is shown that the value of $G$ can be related to the
well-known concurrence $c$ \cite{11} with the relation
\begin{equation}
G=c^{2}(c^{2}+2)\text{.}
\end{equation}%
$G>0$ implies the state is entangled. Therefore, entanglement can be
determined by measuring $G$. For mixed states, Eq. $(4)$ does not hold. As a
substitution, Ref. \cite{Kothe} gives a relation%
\begin{equation}
c^{2}(c^{2}+2)\leq G\leq 2c^{2}+1\text{.}
\end{equation}%
They explain the bounds of $G$\ but do not give strict algebraical proof. In
general, $G$ is an entanglement witness for mixed states.

The experiment set-up is shown in Fig. $1$. A $0.59$ $mm$ thick $\beta $%
-barium borate (BBO) crystal arranged in the Kwiat type configuration \cite%
{8} is pumped by a $351.1$ $nm$ laser beam produced by an Ar$^{\text{+}}$
laser. Through the spontaneous parametric down-conversion (SPDC) process, a
non-maximally entangled state $a|HH\rangle +b|VV\rangle $ ( H and V
represent horizontal and vertical polarization of the photons respectively)
is produced, where the real numbers $a$ and $b$ can be determined by the
polarization of the pump beam and the normalization condition $a^{2}+b^{2}=1$%
.

The measurement of $C$ is easy to implement because%
\begin{equation}
C(\sigma _{i},\sigma _{j})=\langle \sigma _{i}^{A}\otimes \sigma
_{j}^{B}\rangle -\langle \sigma _{i}^{A}\otimes I^{B}\rangle \langle
I^{A}\otimes \sigma _{j}^{B}\rangle \text{,}
\end{equation}%
the value of $C$ just needs combination of coincidence rates of proper
projection measurements. We take $C(\sigma _{1},\sigma _{2})$ for example:

\begin{eqnarray}
C(\sigma _{1},\sigma _{2}) &=&\langle \sigma _{1}^{A}\otimes \sigma
_{2}^{B}\rangle -\langle \sigma _{1}^{A}\otimes I^{B}\rangle \langle
I^{A}\otimes \sigma _{2}^{B}\rangle =  \notag \\
&&\langle |+R\rangle \langle +R|-|+L\rangle \langle +L|-|-R\rangle \langle
-R|+|-L\rangle \langle -L|\rangle  \notag \\
&&-\langle |++\rangle \langle ++|+|+-\rangle \langle +-|-|-+\rangle \langle
-+|-|--\rangle \langle --|\rangle  \notag \\
&&\ast \langle |RR\rangle \langle RR|-|RL\rangle \langle RL|+|LR\rangle
\langle LR|-|LL\rangle \langle LL|\rangle \text{,}
\end{eqnarray}%
where $|\pm \rangle =\frac{1}{\sqrt{2}}(|H\rangle \pm |V\rangle )$ and$\
|R\rangle =$ $\frac{1}{\sqrt{2}}(|H\rangle +i|V\rangle )$ ($|L\rangle =$ $%
\frac{1}{\sqrt{2}}(|H\rangle -i|V\rangle ))$ are the eigenvectors of $\sigma
_{1}$ and $\sigma _{2}$. Then according to Eq. $(3)$, we need to measure the
coincidence rates between every two bases of the set \{$|H\rangle $, $%
|V\rangle $, $|+\rangle $, $|-\rangle $, $|R\rangle $, $|L\rangle $\}\ to
obtain the value of $G$.\ The total number of measurements involved are $36$%
. The number of the usual two-qubit tomography to reconstruct the density
matrix is $16$. The advantage of the measure $G$ is that the entanglement
can be directly obtained from the measurement. On the other hand, we can
estimate the concurrence from the density matrix reconstructed from
tomography.

In order to utilize G as the measure of entanglement, first it is necessary
to demonstrate ILUT of $G$ for two-qubit states. In experiment, we choose
the pure state as $|HV\rangle -|VH\rangle $, $0.91|HH\rangle +0.41|VV\rangle
$, and $0.99|HH\rangle +0.12|VV\rangle $, respectively. For the second and
third groups of data in Table I, two half wave plates (HWP) with the angles
set to $45^{\circ }$\ and $0^{\circ }$\ respectively, are used as the local
unitary transformation. For the other groups of data,\ a quarter wave plate
(QWP) with the angle set to $45^{\circ }$ is used as the local unitary
transformation. From the first three groups of data in Table I, we can see $%
G $ remains invariant under the transformation within the errors. To
demonstrate ILUT of $G$ for mixed states,\ we let entangled states $%
a|HH\rangle +b|VV\rangle $\ pass through\ two same phase-damping channels in
\{$|H\rangle -|V\rangle $,$|H\rangle +|V\rangle $\} basis with the
corresponding superoperators \{$\sqrt{1-p}I$, $\sqrt{p}\sigma _{1}$\}$%
_{A}\otimes $\{$\sqrt{1-p}I$, $\sqrt{p}\sigma _{1}$\}$_{B}$ for the next two
groups of data in Table I and two same phase-damping channels in \{$%
|H\rangle $, $|V\rangle $\} basis with the corresponding superoperators \{$%
\sqrt{1-p}I$, $\sqrt{p}\sigma _{3}$\}$_{A}\otimes $\{$\sqrt{1-p}I$, $\sqrt{p}%
\sigma _{3}$\}$_{B}$\ for the last groups of data\ where $p$ is
connected with the thickness of the quartz and the bandwidth of the
interference filter \cite{quartz}. From Table I, we can see the
values of $G$ are invariant under local unitary transformation
within experimental errors for both pure and mixed states (for the
better understanding of Table I, we list the typical data in Table
II). Therefore, $G$ can be used as meaure of entanglement.
\begin{table}[tbp]
\label{tabone}
\begin{equation*}
\begin{tabular}{p{0.48in}p{0.48in}p{0.48in}p{0.48in}}
\hline\hline
$\quad G$ & $\delta G$ & $G^{\prime }$ & $\delta G^{\prime }$ \\ \hline\hline
$2.841$ & $0.034$ & $2.836$ & $0.034$ \\ \hline\hline
$1.376$ & $0.022$ & $1.419$ & $0.024$ \\ \hline\hline
$0.084$ & $0.006$ & $0.083$ & $0.006$ \\ \hline\hline
$2.831$ & $0.018$ & $2.859$ & $0.018$ \\ \hline\hline
$2.303$ & $0.016$ & $2.324$ & $0.016$ \\ \hline\hline
$0.606$ & $0.009$ & $0.598$ & $0.008$ \\ \hline\hline
\end{tabular}%
\ \
\end{equation*}%
\caption{Demonstration of the invariance of $G$ under local unitary
transformations\emph{.\emph{\ $G^{\prime }$ denotes the values of G after
the local transformation. The first three groups of data are corresponding
to pure states, while the other three ones are for mixed states.}}}
\end{table}

In the process of entanglement\ measurement of pure two-qubit states, we can
obtain the values of $G$ of the experiment-prepared states based on Eq. $(3)$%
. On the other hand, we can obtain the values of concurrence $c$\ according
to these pure states which are denoted by c$1$ in Fig. $2$. We can also
obtain the values of concurrence $c$\ according to the reconstructed density
matrix which are shown in in Fig. $2$\ represented\ with c$2$.\ In Fig. $2$%
(a), the prepared state is $|\varphi _{0}\rangle =\cos 2\theta |HH\rangle
+\sin 2\theta |VV\rangle $, where $\theta $\ (the horizontal axis) is the
angle between the optical axis of the half wave plate (HWP) in pump light
path and the vertical axis. The vertical axis denotes the values of $G$ and $%
c$. We can see that the dots deriving from the experiment data agree\ with
the theoretical\ curves of $G$ and $c$\ plotted according to Eq. $(3)$ and $%
(4)$, respectively. However, the values of concurrence $c$ obtained through
the measurement of G are always higher than those obtained by\ tomography.
This result indicates that the errors arising from the reconstruction in
tomography have an unignorable impact on the experimental measurement of
concurrence since the maximally\ entanglement state produced has a
visibility of $97.8\%$. Therefore, it is more accurate to adopt $G$ as the
measure of entanglement of\ pure states.\ We obtain similar results In Fig. $%
2$(b) for another series of pure states $\cos 2\theta |HV\rangle -\sin
2\theta |VH\rangle $.

\begin{tabular}{||l||l||l||l||l||l||}
\hline\hline
$C(HH)=26$ & $C(HV)=5005$ & $C(HD)=2162$ & $C(HR)=2477$ & $C(VH)=4881$ & $%
C(VV)=16$ \\ \hline\hline
$C(VD)=2558$ & $C(VR)=2416$ & $C(DH)=2738$ & $C(DV)=2303$ & $C(DD)=106$ & $%
C(DR)=2957$ \\ \hline\hline
$C(RH)=2359$ & $C(RV)=2446$ & $C(RD)=2040$ & $C(RR)=96$ & $C(H-)=2939$ & $%
C(HL)=2527$ \\ \hline\hline
$C(V-)=2241$ & $C(VL)=2338$ & $C(D-)=4947$ & $C(DL)=2148$ & $C(-H)=2168$ & $%
C(-V)=2657$ \\ \hline\hline
$C(-D)=4725$ & $C(--)=99$ & $C(-L)=2611$ & $C(-R)=2116$ & $C(LH)=2445$ & $%
C(LV)=2481$ \\ \hline\hline
$C(LD)=2693$ & $C(L-)=2243$ & $C(LL)=75$ & $C(LR)=4853$ & $C(R-)=2796$ & $%
C(RL)=4727$ \\ \hline\hline
\end{tabular}

\begin{tabular}{||l||l||l||l||l||l||}
\hline\hline
$C(HH)=2427$ & $C(HV)=2734$ & $C(HD)=1256$ & $C(HR)=418$ & $C(VH)=2529$ & $%
C(VV)=2275$ \\ \hline\hline
$C(VD)=3564$ & $C(VR)=4605$ & $C(DH)=2086$ & $C(DV)=3155$ & $C(DD)=469$ & $%
C(DR)=3774$ \\ \hline\hline
$C(RH)=4781$ & $C(RV)=54$ & $C(RD)=1803$ & $C(RR)=2397$ & $C(H-)=3763$ & $%
C(HL)=4592$ \\ \hline\hline
$C(V-)=1389$ & $C(VL)=283$ & $C(D-)=4727$ & $C(DL)=1376$ & $C(-H)=2723$ & $%
C(-V)=1995$ \\ \hline\hline
$C(-D)=4378$ & $C(--)=365$ & $C(-L)=3416$ & $C(-R)=1276$ & $C(LH)=45$ & $%
C(LV)=4882$ \\ \hline\hline
$C(LD)=2894$ & $C(L-)=2046$ & $C(LL)=2362$ & $C(LR)=2552$ & $C(R-)=2917$ & $%
C(RL)=2290$ \\ \hline\hline
\end{tabular}

Up to now, all our discussions are confined to local observables, but they
have some disadvantages: LURs can only be constructed to characterize
separable states and its generalization to multipartite and high-dimensional
systems is not very clear. Ref. \cite{Guhne} generalizes LURs to
multipartite and high-dimensional systems utilizing nonlocal observables,
which can overcome these disadvantages. As a basic and important
proposition, two uncertainty relations using nonlocal observables is
considered: For an entangled state $|\psi _{1}\rangle =a|00\rangle
+b|11\rangle $(for simplicity, $a\geq b$), there exist $M_{i}$ such that for
$|\psi _{1}\rangle $,

\begin{equation}
K=\sum_{i}\delta ^{2}(M_{i})_{|\psi _{1}\rangle \langle \psi _{1}|}=0
\end{equation}%
holds, while for separable states%
\begin{equation}
K=\sum_{i}\delta ^{2}(M_{i})\geq 2a^{2}b^{2}
\end{equation}%
is fulfilled. Here $M_{i}=|\psi _{i}\rangle \langle \psi _{i}|$, $i=1,...,4$%
, and $|\psi _{2}\rangle =a|01\rangle +b|10\rangle $, $|\psi _{3}\rangle
=-a|10\rangle +b|01\rangle $, $|\psi _{4}\rangle =b|00\rangle -a|11\rangle $%
. So the violation of\ inequality $(9)$ can be used for entanglement
verification.\ It is also simple to implement in experiment. For
experimental measurement of $K$, the operator $M_{i}$ can be decomposed into
local operators, i.e., written into a sum of projectors onto product vectors
\cite{5}. The key two terms $|00\rangle \langle 11|+|11\rangle \langle 00|$
and $|01\rangle \langle 10|+|10\rangle \langle 01|$\ can be decomposed:

\begin{equation}
|00\rangle \langle 11|+|11\rangle \langle 00|=|RL\rangle \langle
RL|+|LR\rangle \langle LR|+|++\rangle \langle ++|+|--\rangle \langle --|-I
\end{equation}%
and

\begin{equation}
|01\rangle \langle 10|+|10\rangle \langle 01|=|++\rangle \langle
++|+|--\rangle \langle --|-|RL\rangle \langle RL|-|LR\rangle \langle LR|%
\text{,}
\end{equation}%
where $I$ is the $4\times 4$ identity matrix. So $K$ can be experimentally
measured with two-photon coincidence. From Eqs. $(8)$, $(10)$ and $(11)$,
the number of measurement of $K$ is $10$ \cite{10}. However, the number of
measurement of $K$ according to tomography is $16$. So it is more convenient
to use the nonlocal observables to verify entanglement. However, for
nonlocal observables, they can not be measured directly\ and\ the local
decomposition is needed compared with local observables.

To demonstrate entanglement verification using Eqs. $(8)$ and $(9)$, we
choose the entangled state $|\varphi _{0}\rangle $ and two product states
\TEXTsymbol{\vert}$\varphi _{1}\rangle =|HH\rangle $, $|\varphi _{2}\rangle
=|VH\rangle $. By means of local decomposition of $M_{i}$,\ the values of $K$
are measured for these states respectively. In Fig. $3$, $Ki$ ($i=0,1,2$)
denotes the values of $K$ for $|\varphi _{i}\rangle $. In this Figure, all
the points about K$1$ and K$2$ are above the curve representing the lower
bound for inequality $(9)$. K$0$ are well below this curve, approaching its
theoretical value zero, which is a good indication of entanglement
verification\ for pure states.

The basic idea to take one $M_{i}$\ as the projector onto the range of state
space, and the other $M_{i}$\ as projectors onto the basis of the kernel,
can also be generalized to other cases. For example, it can be generalized
to verify entanglement of arbitrary bipartite $N\times M$\ system. For
multipartite systems, the investigation of nonlocal observables can play an
important role in verifying multipartite entangled states \cite{Guhne}.

In a summary, we experimentally test ILUT of the new measure$\ G$ for both
pure and mixed states. By means of ILUT, we demonstrate entanglement
quantification using $G$\ for pure states with the forms of $\cos 2\theta
|HH\rangle +\sin 2\theta |VV\rangle $ and $\cos 2\theta |HV\rangle -\sin
2\theta |VH\rangle $. Since any two-qubit pure states can be decomposed into
these forms under Schmidt decomposition up to certain local unitary
transformations, the method can be generalized to all two-qubit pure states.
The measure can be generalized to high-dimensional systems which would keep
the properties like ILUT and may be useful to detect and quantify
entanglement. We also generalize to nonlocal observables which can overcome
the disadvantages brought by LURs. Experimentally we demonstrate
Entanglement verification using\ nonlocal uncertainty relations Eq. $(8)$
and inequality $(9)$ for pure states, which can be subsequently generalized
to arbitrary two-qubit pure states. The investigation of nonlocal
observables can play an important role in verifying multipartite entangled
states and distinguishing between different classes of true tripartite
entanglement for qubits. \

\begin{center}
\textbf{ACKNOWLEDGMENTS}
\end{center}

The authors would like to thank D. Cavalcanti for helpful comment. This work
was funded by the National Fundamental Research Program, National Natural
Science Foundation of China (Grant No.10674127 and 60121503), Innovation
Funds from Chinese Academy of Sciences, and Program for New Century
Excellent Talents in University.

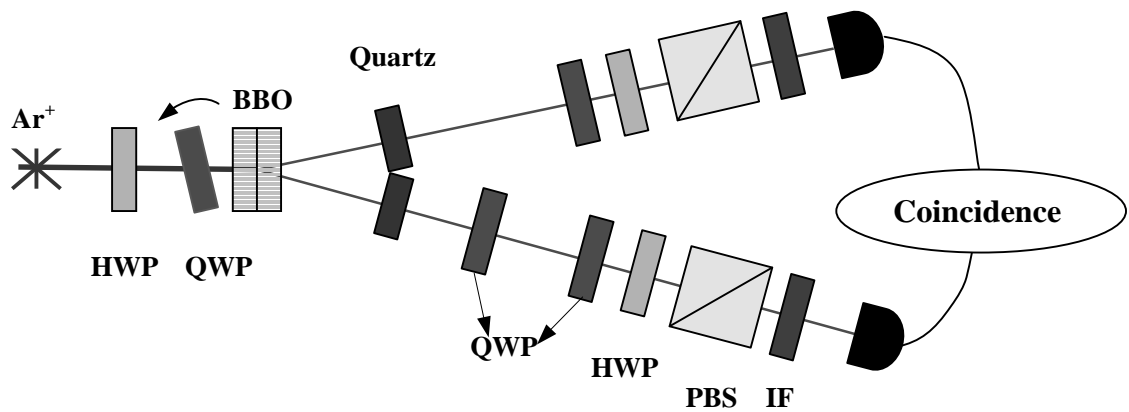
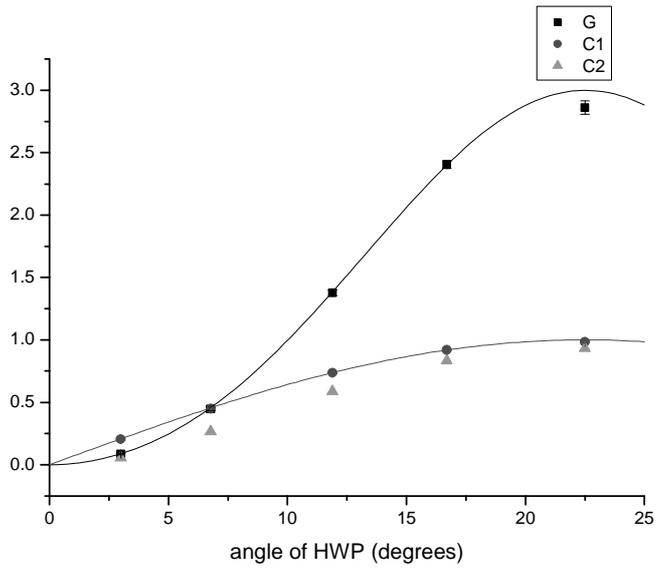
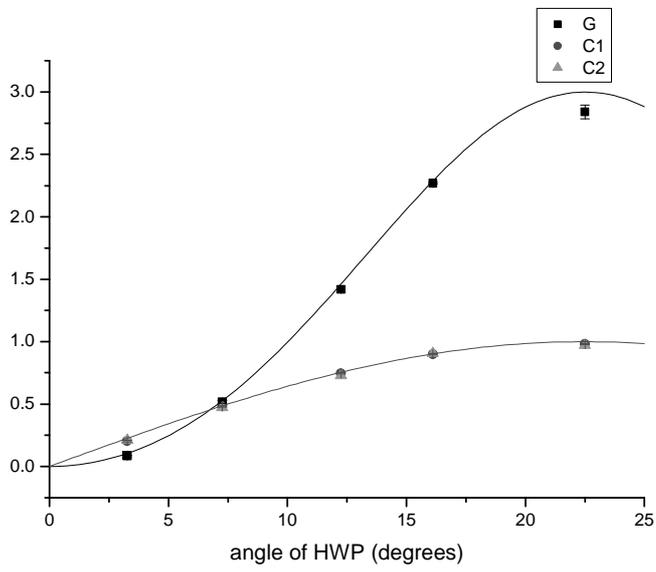
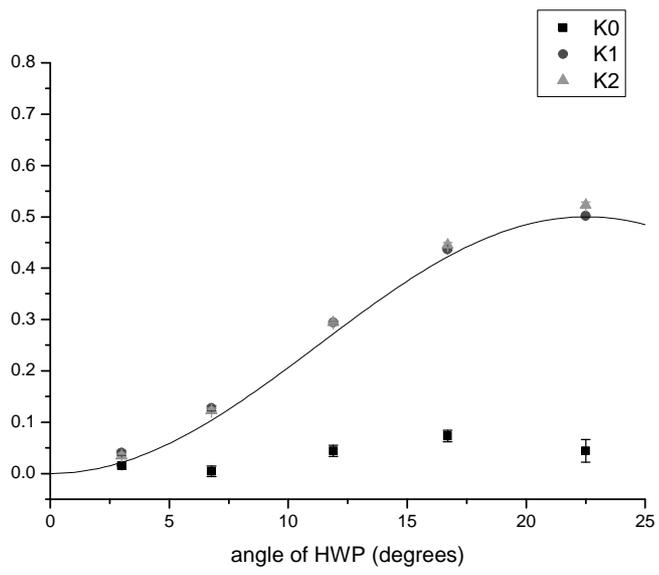
\end{document}